\documentclass[sigconf]{acmart}

\usepackage{booktabs} 
\usepackage{color}
\usepackage{listings}
\usepackage[labelformat=simple]{caption}
\usepackage{graphicx}
\usepackage{subcaption}
\usepackage{nopageno}

\usepackage{xcolor}
\usepackage{soul}

\graphicspath{ {Figs/} }

\begin{document}

\copyrightyear{2018}
\acmYear{2018}
\setcopyright{acmcopyright}
\acmConference[WSDM 2018]{WSDM 2018: The Eleventh ACM International Conference on Web Search and Data Mining }{February 5--9, 2018}{Marina Del Rey, CA, USA}
\acmPrice{15.00}
\acmDOI{10.1145/3159652.3159684}
\acmISBN{978-1-4503-5581-0/18/02}

\title{Can you Trust the Trend? Discovering Simpson's \\ Paradoxes in Social Data}


\author{Nazanin Alipourfard}
\authornote{N. Alipourfard and P. Fennell contributed equally to this work.}
\affiliation{%
  \institution{USC/ISI}
  \streetaddress{4676 Admiralty Way, Marina Del Rey}
  \city{Los Angeles}
  \state{California}
  \postcode{90292}
}
\email{nazanina@isi.edu}

\author{Peter G. Fennell}
\affiliation{%
  \institution{USC/ISI}
  \streetaddress{4676 Admiralty Way, Marina Del Rey}
  \city{Los Angeles}
  \state{California}
  \postcode{90292}
}
\email{pfennell@isi.edu}


\author{Kristina Lerman}
\affiliation{%
  \institution{USC/ISI}
  \streetaddress{4676 Admiralty Way, Marina Del Rey}
  \city{Los Angeles}
  \state{California}
  \postcode{90292}
}
\email{lerman@isi.edu}

\renewcommand{\shortauthors}{Alipourfard, Fennell et al.}

\begin{abstract}
We investigate how Simpson's paradox affects analysis of trends in social data. According to the paradox, the trends observed in data that has been aggregated over an entire population may be different from, and even opposite to, those of the underlying subgroups. Failure to take this effect into account can lead analysis to wrong conclusions. We present a statistical method to automatically identify Simpson's paradox in data by comparing statistical trends in the aggregate data to those in the disaggregated subgroups.
We apply the approach to data from Stack Exchange, a popular question-answering platform, to analyze factors affecting answerer performance, specifically, the likelihood that an answer written by a user will be accepted by the asker as the best answer to his or her question. Our analysis confirms a known Simpson's paradox and identifies several new instances. These paradoxes provide novel insights into user behavior on Stack Exchange.

\end{abstract}

%
%



\maketitle

\section{Introduction}
%

Digital traces of human activity have exposed social behavior to web and data mining algorithms. Researchers have analyzed the growing social data to understand, among other things, how information spreads in social networks~\cite{Romero11www}, and the factors affecting user engagement~\cite{Barbosa2016} and performance~\cite{Singer2016plosone} in online platforms. Yet social data analysis presents multi-faceted challenges that existing data mining approaches are not well-equipped to handle.
Real-world data is noisy and sparse. To uncover hidden patterns, scientists aggregate data over the population, which tends to improve statistics and signal-to-noise ratio. However, real-world data is also heterogeneous, i.e., composed of subgroups that vary widely in size and behavior. This heterogeneity can severely bias analysis of social data. 

One example of such bias is Simpson's paradox. According to the paradox, 
an association observed in data that has been aggregated over an entire population may be quite different from, and even opposite to, those of the underlying subgroups. Failure to take this effect into account can distort conclusions drawn from data.
One of the better known examples of Simpson's paradox comes from a study of gender bias in graduate admissions~\cite{Bickel1975}. In the aggregate admissions data there appears to be a statistically significant bias against women: a smaller fraction of female applicants is admitted for graduate studies. However, when admissions data is disaggregated by department, women have parity and even a slight advantage over men in some departments. The paradox arises because departments preferred by female applicants have lower admissions rates for both genders.

Simpson's paradox also affects analysis of trends. When measuring how an outcome changes as a function of an independent variable, the characteristics of the population over which the trend is measured may change as a function of the independent variable due to survivor bias. As a result, the data may appear to exhibit a trend, which disappears or reverses when the data is disaggregated. 
For example, it may appear that repeated exposures to information or hashtags on social media make an individual less likely to share it with his or her followers~\cite{Romero11www}. In fact, the opposite is true: the more people are exposed to information, the more likely they are to share it with followers~\cite{Lerman2016futureinternet}. However, those who follow many others (and are likely to be exposed to a meme or a hashtag multiple times) are less responsive overall, due to the high volume of information they receive~\cite{Hodas12socialcom}. Their reduced susceptibility to exposures biases the aggregate response, leading to wrong conclusions about behavior. Once disaggregated based on the volume of information received, a clearer pattern of exposure response emerges, one that is more predictive of the actual response~\cite{Hodas14srep}.
However, despite accumulating evidence that Simpson's paradox 
affects inference of trends in social and behavioral data~\cite{Singer2016plosone,Barbosa2016,KootiA2017www,Agarwal2017quitting}, researchers do not routinely test for it in their studies.

We describe a method to identify Simpson's paradoxes in the analysis of trends in social data. 
Our statistical approach finds pairs of variables---that we call \emph{Simpson's pairs}---such that a trend in some outcome as a function of the first \emph{independent} variable observed in the aggregate data  disappears or reverses itself when the data is disaggregated into distinct subgroups on the second \emph{conditioning} variable.
We perform mathematical analysis, which identifies two necessary conditions for the paradox to occur: (1) the independent and conditioning variables are correlated and (2) the value of the outcome variable differs within conditioning subgroups.
%

We apply the proposed approach to data collected from Stack Exchange, a popular question-answering platform. Specifically, we analyze factors affecting answerer performance, measured as the likelihood the answer provided by the answerer will be accepted by the asker as the best answer to his or her question. We construct a variety of features to describe answers and answerers, and study how the outcome---in this case answer acceptance---depends on these features. We compare results of statistical trends found in aggregated and disaggregated data to identify Simpsons's pairs. Our analysis discovers several cases of Simpson's paradox in Stack Exchange data. In addition to a known effect that describes deterioration of answerer performance over the course of a session, our method identifies several new cases of Simpson's paradox. These paradoxes yield new insights into answerer performance on Stack Exchange. For example, creating a new feature to describe answerers, which combines variables of a Simpson's pair, leads to a more robust proxy of answerer performance.

Presence of a Simpson's paradox in social data can indicate interesting patterns~\cite{fabris2000discovering}, including important behavioral differences. The paradox suggests that subgroups within the population under study systematically differ in their behavior, and these differences are large enough to affect aggregate trends. In such cases the trends discovered in disaggregated data are more likely to describe---and predict---individual behavior than the trends found in aggregated data. Thus, to build more robust models of behavior, data scientists need to identify the subgroups within their data 
or risk drawing wrong conclusions.
The method presented in this paper provides a simple framework for identifying such interesting subgroups by systematically searching for Simpson's paradox in trend data.

The rest of the paper is organized as follows. First, we present background, including multiple observations of Simpson's paradox in a variety of disciplies (Sec.~\ref{sec:related}). Then we describe our methodology for detecting Simpson's paradox by identifying covariates in data, and analyze the paradox mathematically to gain more insight into its origins (Sec.~\ref{sec:methods}). Finally, we apply our method to real-world data from the question-answering site Stack Exchange, and demonstrate its ability to automatically identify novel cases of Simpson's paradox (Sec.~\ref{sec:results}). We conclude with the discussion of implications.

\section{Background and Related Work}
\label{sec:related}
%
%
%
%
%

%
%
%

The goal of data analysis is to identify important associations between features, or variables, in data. 
However, hidden correlations between variables can lead analysis to wrong conclusions. One important manifestation of this effect is Simpson's paradox, according to which \emph{an association that appears in different subgroups of data may disappear, and even reverse itself, when data is aggregated across subgroups}.
Instances of the paradox have been documented across a variety of disciplines, including demographics, economics, political science, and clinical research, and it has been argued that the presence of Simpson's paradox implies that interesting patterns exist in data~\cite{fabris2000discovering}.
A notorious example of Simpson's paradox  arose during a gender bias lawsuit against UC Berkeley. Analysis of graduate school admissions data revealed a statistically significant bias against women. 
However, the pattern of discrimination observed in this aggregate data disappeared when admissions data was disaggregated by department.
Bickel et al.~\cite{Bickel1975} attributed this effect to Simpson's paradox. They argued that the subtle correlations between the popularity of departments among the genders and their selectivity resulted in women applying to departments that were hardest to get into, which skewed analysis.

Simpson's paradox must also be considered in the analysis of trends. Vaupel and Yashin~\cite{Vaupel85heterogeneity} give a compelling illustration of how survivor bias can shift the composition of data, distorting the conclusions drawn from it. Analysis of recidivism among convicts released from prison shows that the rate at which they return to prison declines over time. From this, policy makers may conclude that age has a pacifying effect on crime: older convicts are less likely to commit crimes. In reality, this conclusion is false. 
Instead, we can think of the population of ex-convicts as composed of two subgroups with constant, but very different recidivism rates. The first subgroup, let's call them ``reformed,'' will never commit a crime once released from prison. The other subgroup, the ``incorrigibles,'' will always commit a crime. Over time, as ``incorrigibles'' commit offenses and return to prison, there are fewer of them left in the population. The survivor bias changes the composition of the population under study, creating an illusion of an overall decline in recidivism rates. As Vaupel and Yashin warn, ``unsuspecting researchers who are not wary of heterogeneity's ruses may fallaciously assume that observed patterns for the population as a whole also hold on the sub-population or individual level.'' Their paper gives numerous other examples of such \emph{ecological fallacies}.

Similar illusions crop up in many studies of social behavior. For example, when examining how social media users respond to information from their friends (other users that they follow), it may appear that if \emph{more} of a user's friends use a hashtag then the user will be \emph{less} likely to use it himself or herself~\cite{Romero11www}. Similarly, the more friends share some information, the less likely the user is to share it with his or her followers~\cite{Versteeg11icwsm}. From this, one may conclude the additional exposures to information in a sense ``innoculate'' the user and act to suppress the sharing of information. In fact, this is not the case, and instead, additional exposures monotonically increase the user's likelihood to share information with followers~\cite{Lerman2016futureinternet}. However, those users who follow many others, and are likely to be exposed to information or a hashtag multiple times, are less responsive overall, because they are overloaded with information they receive from all the friends they follow~\cite{Hodas12socialcom}. Calculating response as a function of the number of exposures in the aggregate data falls prey to survivor bias: the more responsive users (with fewer friends) quickly drop out of the average (since they are generally exposed fewer times), leaving the highly connected, but less responsive, users behind.  The reduced susceptibility of these highly connected users biases aggregate response, leading to wrong conclusions about individual behavior. Once data is disaggregated based on the volume of information individuals receive, a clearer pattern of response emerges, one that is more predictive of behavior~\cite{Hodas14srep}. Multiple examples of Simpson's paradox have been identified in empirical studies of online behavior. A study~\cite{Barbosa2016} of Reddit found that while it may appear that average comment length on decreases over any fixed period of time, when data is disaggregated into groups based on the year user joined Reddit, comment length within each group increases during the same time period. It is only because users who joined early tend to write longer comments that the Simpson's paradox appears.

Data heterogeneity also impacts statistical analysis of data~\cite{estes1956problem} and causal inference~\cite{Xie2013population}. However, no general framework to recognize and mitigate Simpson's paradox exist. Current methods require that the structure of data be explicitly specified~\cite{Bareinboim2016causal} or at best be guided by subject matter knowledge~\cite{Hernan2011}.


Despite accumulating evidence that Simpson's paradox affects inference from data~\cite{Xie2013population,estes1956problem}, scientists do not routinely test for the presence of this paradox in heterogeneous data. 
Our work addresses this knowledge gap by proposing a statistical method to systematically uncover instances of Simpson's paradox in data.

\section{Methods}
\label{sec:methods}

We propose a method to systematically uncover Simpson's paradox for trends in data. We denote as $Y$ the dependent variable in the data set, i.e., an outcome being measured, and as $\mathbf{X} = \{X_1, X_2,\dots,X_m\}$ the set of $m$ independent variables or features. The goal of the method is to identify pairs of variables $(X_p, X_c)$---\emph{Simpson's pairs}---such that 
a trend in $Y$ as a function of $X_p$ disappears or reverses when the data is disaggregated by conditioning on $X_c$. 
More specifically, our method searches for pairs of variables $(X_p, X_c)$ such that
\begin{align}
	&\frac{d}{dx_p}\mathbb{E}[Y|X_p = x_p] > 0\;\;\;\; \forall x_p, \label{eq:trend_sp1} \\
	&\frac{d}{dx_p}\mathbb{E}[Y|X_p = x_p, X_c = x_c] \leq 0 \;\;\;\; \forall x_p, x_c.
	\label{eq:trend_sp2}
\end{align}
and vice versa (i.e., $d\mathbb{E}[Y|X_p = x_p]/dx_p < 0$, $d\mathbb{E}[Y|X_p = x_p, X_c = x_c]/dx_p \geq 0$). Equations~\eqref{eq:trend_sp1} and \eqref{eq:trend_sp2} hold if the expected value of $Y$ is a monotonically increasing (or decreasing) function of $X_p$ alone, but conditioned on $X_c$ is a monotonically decreasing (resp. increasing) function of $X_p$, or is constant.

\subsection{Finding Simpson's Pairs}
With this goal in mind, we employ linear models to quantify the relationship between $Y$, an independent variable $X_p$, and a conditioning variable $X_c$ upon which the data is disaggregated. Firstly, on the aggregate level, we model the relationship between $Y$ and $X_p$ as a linear model of the form
\begin{equation}
	\mathbb{E}[Y|X_p = x_p] = f_p(\alpha + \beta x_p),
	\label{eq:f_p}
\end{equation}
where $f_p(\alpha + \beta x_p)$ is a monotonically increasing function of its argument $\alpha + \beta x_p$. The parameter $\alpha$ in Eq.~\eqref{eq:f_p} is the intercept of the regression function, while the trend parameter $\beta $ quantifies the effect of $X_p$ on $Y$. Secondly, for the disaggregation, we fit linear models of the form of Eq.~\eqref{eq:f_p} but with different values of the parameters $\alpha$ and $\beta $ depending on the value of $X_c$:	
\begin{equation}
	\mathbb{E}[Y|X_p = x_p, X_c = x_c] = f_{p,c}(\alpha(x_c) + \beta (x_c)x_p),
	\label{eq:f_pc}
\end{equation}


When fitting linear models $f(\alpha + \beta X)$ we have not only a fitted trend parameter $\beta$ but also a $p$-value which gives the probability of finding an intercept $\beta$ at least as extreme as the fitted value under the null hypothesis $H_0:\beta = 0$. From this, we have three possibilities:
\begin{itemize}
  \item $\beta$ is not statistically different from zero (sgn$(\beta)$ = 0), 
  \item $\beta$ is statistically different from zero and positive (sgn$(\beta)$ = 1), 
  \item $\beta$ is statistically different from zero and negative (sgn$(\beta)$ = -1).
\end{itemize}
This mechanism allows us to test for Simpson's paradox by comparing the sign of $\beta$ from the aggregated fit (Eq.~\eqref{eq:f_p}) with the signs of the $\beta$'s from the disaggregated fits (Eq.~\eqref{eq:f_pc}). 
{Although} Eqs.~\eqref{eq:f_p} and \eqref{eq:f_pc} 
{state that the signs from the disaggregated curves should all be different from the aggregated curve, in practice this is too strict, especially as human behavioral data is noisy. Thus, we compare the sign of the fit to aggregated data to the simple average of the signs of fits to disaggregated data--- if the signs are different then we have uncovered an instance of Simpson's paradox}. The summary of the algorithm is the following:


\lstset{language=Python,
  showspaces=false,
  showtabs=false,
  breaklines=true,
  numbers=left,
  stepnumber=1,
  frame=single,
  showstringspaces=false,
  breakatwhitespace=true,
  escapeinside={(*@}{@*)},
  commentstyle=\color{greencomments},
  stringstyle=\color{redstrings},
  basicstyle=\ttfamily
}

\begin{lstlisting}[title=Trend Simpson's Paradox Algorithm][language=Python]
def trend_simpsons_pair(X, Y):
 paradox_pairs = []
 for paradox_var in vars:
  beta, pvalue = trend_analysis(X[paradox_var], Y)
  agg = sgn(beta, pvalue)
  for condition_var in vars:
   if paradox_var != condition_var:
    dagg = []
    for con_gr in bins_of(condition_var):
     beta, pvalue = trend_analysis(X[paradox_var | con_gr], Y)
     dagg.append(sgn(beta, pvalue))
    if agg != sgn(mean(dagg)):
     paradox_pairs.append([paradox_var, condition_var])
 return paradox_pairs

 def sgn(beta, pvalue = 0.0):
  return (0 if (beta == 0 or pvalue > 0.05) else (1 if beta > 0 else -1))
\end{lstlisting}

\paragraph{Data Disaggregation}
A critical step in our method is disaggregating data by conditioning on variable $X_c$. The idea behind disaggregation is to segment data into more homogeneous subgroups of similar elements. For multinomial variables $X_c$, disaggregation step simply involves grouping data by unique values of $X_c$. However, for continuous $X_c$ or discrete variables with large range, this step is more complex. We can bin the elements according to their values of $X_c$, but the decision has to be made how large each bin is, whether bin sizes scale linearly or logarithmically, etc. If the bin is too small, it may not contain enough samples for a statistically significant measurement, but if it is too large, the samples may be too heterogeneous for a robust trend. 
{In our experiments described below, bins of fixed size successfully identify Simpson's pairs, though we realize that more sophisticated binning techniques can allow us to isolate more pairs or reduce the number of false positives.}

\subsection{Mathematical Analysis of the Paradox}
\label{sec:math}
We have presented a mathematical formulation of Simpson's paradox in terms of the derivatives of conditional expectations as given by Eqs.~\eqref{eq:trend_sp1} and \eqref{eq:trend_sp2}, and
we now examine these equations to get a better insight into the origins and causes of this paradox.

The expectation in Eq.~\eqref{eq:trend_sp1} can be related to that of Eq.~\eqref{eq:trend_sp2} as
\begin{eqnarray}
	& & \mathbb{E}[Y|X_p =  x_p] = \label{eq:cond_exp} \\ \nonumber
 & & \int_{X_c}\mathbb{E}[Y|X_p = x_p, X_c = x_c]\Pr(X_c=x_c|X_p=x_p)dx_c,
\end{eqnarray}
and differentiating this expectation w.r.t. $x_p$ allows us to compare the trends of Eqs.~\eqref{eq:trend_sp1} and \eqref{eq:trend_sp2}. The derivative of the right hand side of Eq.~\eqref{eq:cond_exp} with respect to $x_p$ is
\begin{align}
	\int_{X_c}\left(\frac{d}{dx_p}\mathbb{E}[Y|X_p = x_p, X_c = x_c]\right)\Pr(X_c=x_c|X_p=x_p)dx_c + \nonumber \\
	\int_{X_c}\mathbb{E}[Y|X_p = x_p, X_c = x_c]\left(\frac{d}{dx_p}\Pr(X_c=x_c|X_p=x_p)\right)dx_c.
\label{eq:exp_der2}
\end{align}
If $\mathbb{E}[Y|X_p = x_p, X_c = x_c]$ is a non-increasing function of $x_p$---as in Eq.~\eqref{eq:trend_sp2}---then the first integral in Eq.~\eqref{eq:exp_der2} will be non-positive. Thus for $\mathbb{E}[Y|X_p = x_p]$ to be an increasing function of $x_p$, i.e., for Eq.~\eqref{eq:exp_der2} to be positive, the second integral must be positive.

This inequality condition leads to two necessary conditions for the occurrence of Simpson's paradox. The first condition is that
\begin{equation}
	\frac{d}{dx_p}\Pr(X_c=x_c|X_p=x_p) \neq 0,
        \label{eq:SPcond1}
\end{equation}
i.e., the distribution of the conditioning variable $X_c$ is not independent of $X_p$ and so the two variables are correlated. As $X_p$ changes, the distribution of the values of $X_c$ must also change. In the case that the distribution of $X_c$ is independent of $X_p$, then $d\Pr(X_c=x_c|X_p=x_p)/dx = 0$ and so the second integral of Eq.~\eqref{eq:exp_der2} will be zero resulting in no Simpson's paradox.

The second necessary condition for the occurrence of Simpson's paradox is that the expectation of $Y$, conditioned on $X_p$, must not be independent of $X_c$, i.e.,
\begin{equation}
	\mathbb{E}[Y|X_p = x_p, X_c = x_c] \neq \mathbb{E}[Y|X_p = x_p].
        \label{eq:SPcond2}
\end{equation}
For any given value of $X_p$, the expectation of $Y$ must vary as a function of $X_c$. If the condition of Eq.~\eqref{eq:SPcond2} is not met then the second integral in Eq.~\eqref{eq:exp_der2} becomes
\begin{align}
\label{eq:SP_cond2b}
	&\int_{X_c}\mathbb{E}[Y|X_p = x_p]\left(\frac{d}{dx_p}\Pr(X_c=x_c|X_p=x_p)\right)dx_c \\ \nonumber
	&= \mathbb{E}[Y|X_p = x_p]\frac{d}{dx_p}\left(\int_{X_c}\Pr(X_c=x_c|X_p=x_p)dx_c\right) = 0,
\end{align}
and so Simpson's paradox will not occur. 

Thus, this mathematical analysis has given us an insight into causes for Simpson's paradox in data ---correlations between independent variables and the fact that the distribution of the conditioning variable $X_c$ changes at a faster rate with respect to the independent paradox variable $X_p$ than does the expectation of $Y$. This point will be covered in greater detail in the next section.

\section{Results}
\label{sec:results}

We explore our approach using data from the question-answering platform called Stack Exchange. This platform, launched in 2008 to provide a forum for people to ask computer programming questions, grew over the years as a forum asking questions on a variety of technical and non-technical topics.
The premise behind Stack Exchange is simple: any user can ask a question, which others may answer. Users can also \emph{vote} for answers they find helpful, and the asker can \emph{accept} one of the answers as the best answer to the question. In this way, the Stack Exchange community collectively curates knowledge.

\subsection{Stack Exchange Data}
We used anonymized data representing all questions and answers from August 2008 until September 2014.\footnote{\url{https://archive.org/details/stackexchange}} The data includes
9.6M questions, of which approximately half had an accepted answer.
Only the questions that received two or more answers were included~\cite{Burghardt2017myopia}.
Previous studies of Stack Exchange~\cite{Ferrara2017dynamics} and other online platforms~\cite{Agarwal2017quitting,Singer2016plosone}, identified user sessions as an important variable in understanding performance. User actions, i.e., answering questions, can be segmented into \emph{sessions}---periods of activity without a prolonged break. Following \cite{Ferrara2017dynamics} we use 100 minutes as minimum break length to define sessions. A time interval longer than 100 minutes between two answers constitutes the end of one session and the start of a new one. Note that the exact value of this threshold does not change results, but simply merge a small fraction of sessions into longer sessions.

To understand factors affecting user performance on Stack Exchange, we study the relationship between user attributes and the probability that the answer the user produces is accepted by the asker as best answer to his or her question.
To this end, for each answer in the data set, we create a list of features describing the answer, as well as features of describing the user writing the answer:
\begin{description}
  \item[Reputation:] Answerer's {reputation} at the time he or she posted the answer. 
   This score summarizes the user's cumulative contributions to Stack Exchange.
  \item[Number of answers:] Cumulative number of answers written by the user at the time the current answer was posted.
  \item[Tenure:] Age of the user's account (in seconds) at the time the user posted the answer. 
  \item[Percentile:] User's percentile rank based on tenure.
  \item[Time since previous answer:] Time interval (in seconds) since user's previous answer.
  \item[Session length:] The length of the session (in number of answers posted) during which the answer was posted.
  \item[Answer position:] Index of the answer within a session.
  \item[Words:] Number of words in the answer.
  \item[Lines of codes:] Number of lines of codes in the answer.
  \item[URLs:] Number of hyperlinks in the answer.
  \item[Readability:] Answer's Flesch Reading Ease~\cite{Readability} score.
  \end{description}


\subsection{Simpson's Paradoxes on Stack Exchange}

\begin{figure*}[tbh!]
\begin{subfigure}[b]{0.5\textwidth}
  \centering
  \includegraphics[width=\textwidth]{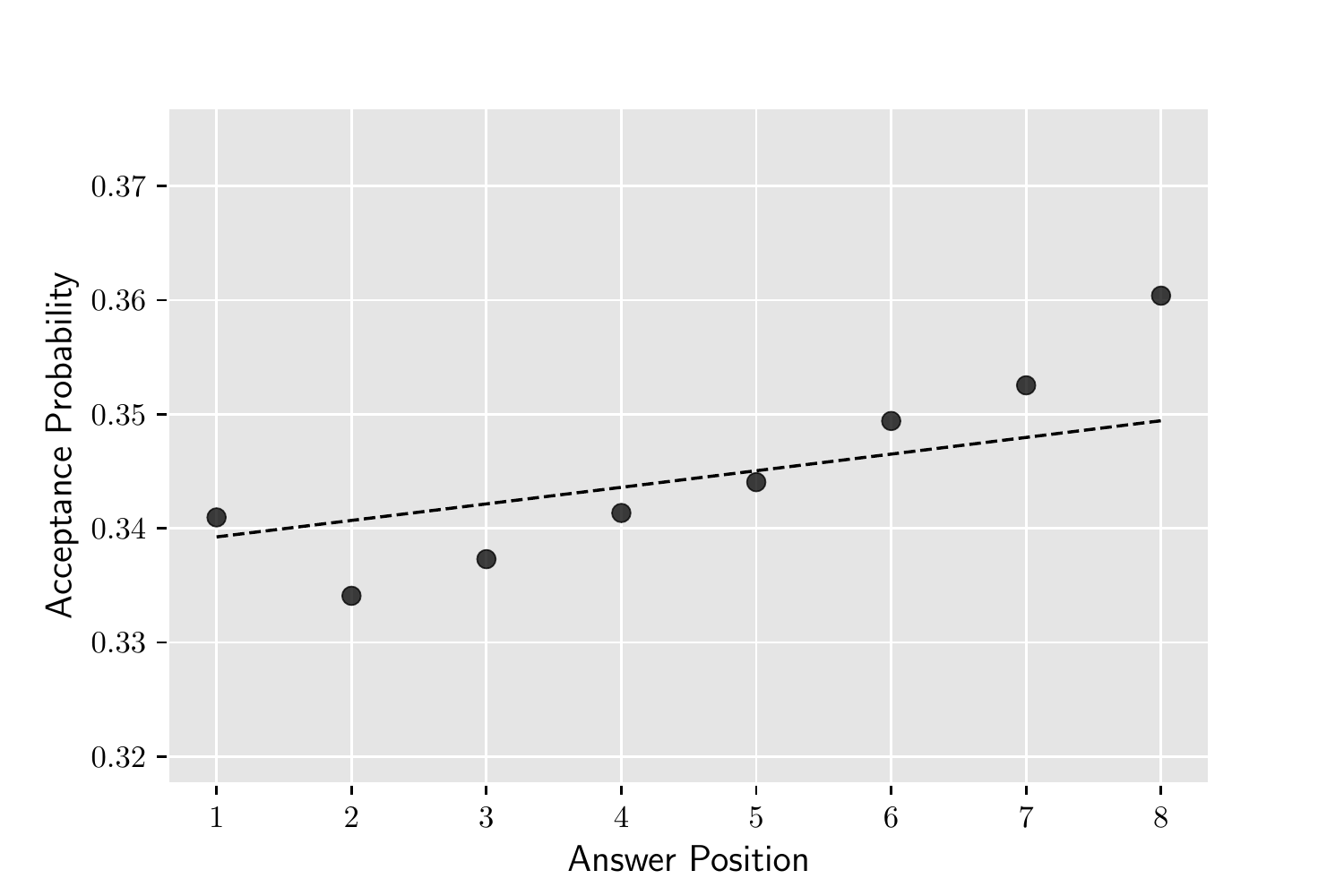}
  \caption{Aggregated Data}
  \label{fig:fig1}
\end{subfigure}%
\begin{subfigure}[b]{0.5\textwidth}
  \centering
  \includegraphics[width=\textwidth]{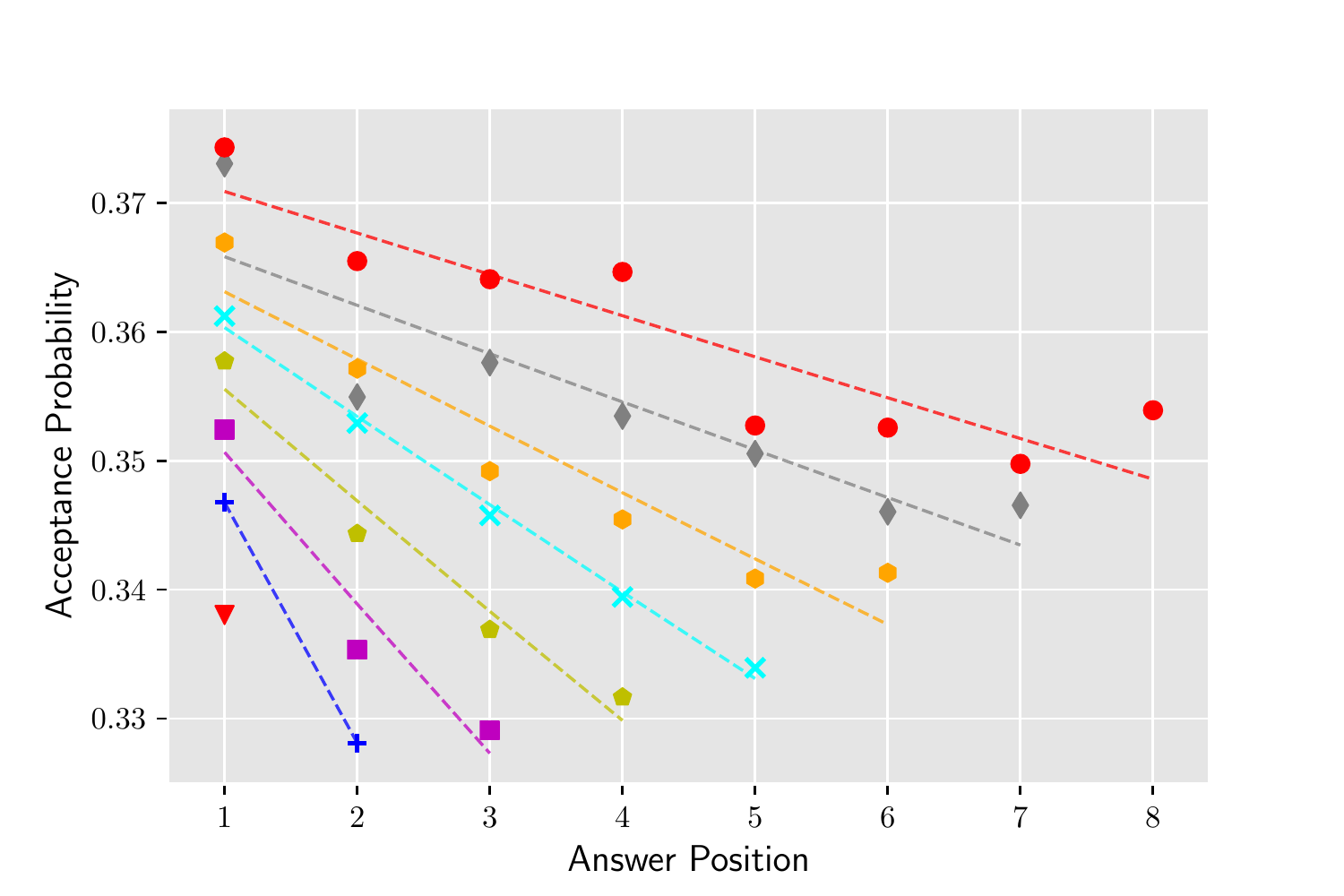}
  \caption{Disaggregated Data}
  \label{fig:fig2}
\end{subfigure}
\caption{
Simpson's paradox in Stack Exchange data. Both plots show the probability an answer is accepted as the best answer to a question as a function of its position within user's activity session. (a) Acceptance probability calculated over aggregated data has an upward trend, suggesting that answers written later in a session are more likely to be accepted as best answers. However, when data is disaggregated by session length (b), the trend reverses. Among answers produced during sessions of the same length (different colors represent different-length sessions), later answers are less likely to be accepted as best answers.
}
\label{fig:session}
\end{figure*}

We apply the method described above to Stack Exchange data. Here, our dependent variable $Y$ is binary, denoting whether or not a specific answer to a question was accepted as the best answer. In this case of binary outcomes we use the logistic regression linear model of the form
\begin{equation}
	f(\alpha + \beta x) = \frac{1}{1+e^{-(\alpha + \beta x)}}.
\end{equation}
The parameters $\alpha$ and $\beta$ are fitted using Maximum likelihood, while test of the null hypothesis $H_0:\beta = 0$ is performed using the Likelihood Ratio Test~\cite{casella2002statistical}.

\begin{table}
  \begin{tabular}{|c|c|c|}
    \toprule
    \textbf{$X_p$: Independent Variable}&\textbf{$X_c$: Conditioning Variable}\\
    \midrule
    Tenure&Number of answers\\
    Session length&Reputation\\
    Answer position&Reputation\\
    Answer position&Session length\\
    Number of answers&Reputation\\
    Time since previous answer&Answer position\\
    Percentile&Number of answers\\
  \bottomrule
\end{tabular}
\caption{Examples of Simpson's paradox in Stack Exchange data. For these variables, the trend in the outcome variable (answer acceptance) as a function of $X_p$ in the aggregate data reverses when the data disaggregated on $X_c$.}   \label{tab:pairs}
\end{table}

The eleven variables in Stack Exchange data, result in 110 possible Simpson's pairs. 
Among these, our method identifies seven as instance of paradox. These are listed in Table~\ref{tab:pairs}.

Our approach reveals that the previously reported finding that acceptance probability decreases with answer position~\cite{Ferrara2017dynamics} is an instance of Simpson's paradox and would not have been observed  had the data not been disaggregated by session length. More interestingly, our approach also identifies previously unknown instances of Simpson's paradox.  We explore these in greater detail below, illustrating how it  can lead to deeper insights into online behavior.

\subsubsection{Answer Position \& Session Length}

We measure session length by the number of answers a user posts before taking an extended break.
Session length was shown to be an important confounding variable in online activity. Analysis of the quality of comments posted on a social news platform Reddit showed that, once disaggregated by the length of session, the quality of comments declines over the course of a session, with each successive comment written by a user becoming shorter, less textually complex, receiving fewer responses and a lower score from others~\cite{Singer2016plosone}. Similarly, each successive answer posted during a session by a user on Stack Exchange is shorter, less well documented with external links and code, and  less likely to be accepted by the asker as the best answer~\cite{Ferrara2017dynamics}. 

Our approach automatically identifies this example as Simpson's paradox, as illustrated in Fig.~\ref{fig:session}. The figure shows average acceptance probability for an answer as a function of its position (or index) within a session. According to Fig.~\ref{fig:fig1}, which reports \emph{aggregate} acceptance probability, answers written later in a session are more likely to be accepted than earlier answers. However, once the same data is \emph{disaggregated by session length}, the trend reverses (Fig.~\ref{fig:fig2}): each successive answer within the same session is less likely to be accepted than the previous answer. For example, for sessions during which five answers were written, the first answer is more likely to be accepted than the second answer, which is more likely to be accepted than the third answer, etc., which is more likely to be accepted than the fifth answer.
The lines in Fig.~\ref{fig:session}  represent fits to data using logistic regression.

This example highlights the necessity to properly disaggregate data to identify the subgroups for analysis. Unless data is disaggregated, wrong conclusions may be drawn, in this case, for example, that user performance improves during a session. 


\subsubsection{Number of Answers \& Reputation}
\begin{figure*}[tbh!]
\begin{subfigure}[b]{0.5\textwidth}
  \centering
  \includegraphics[width=\textwidth]{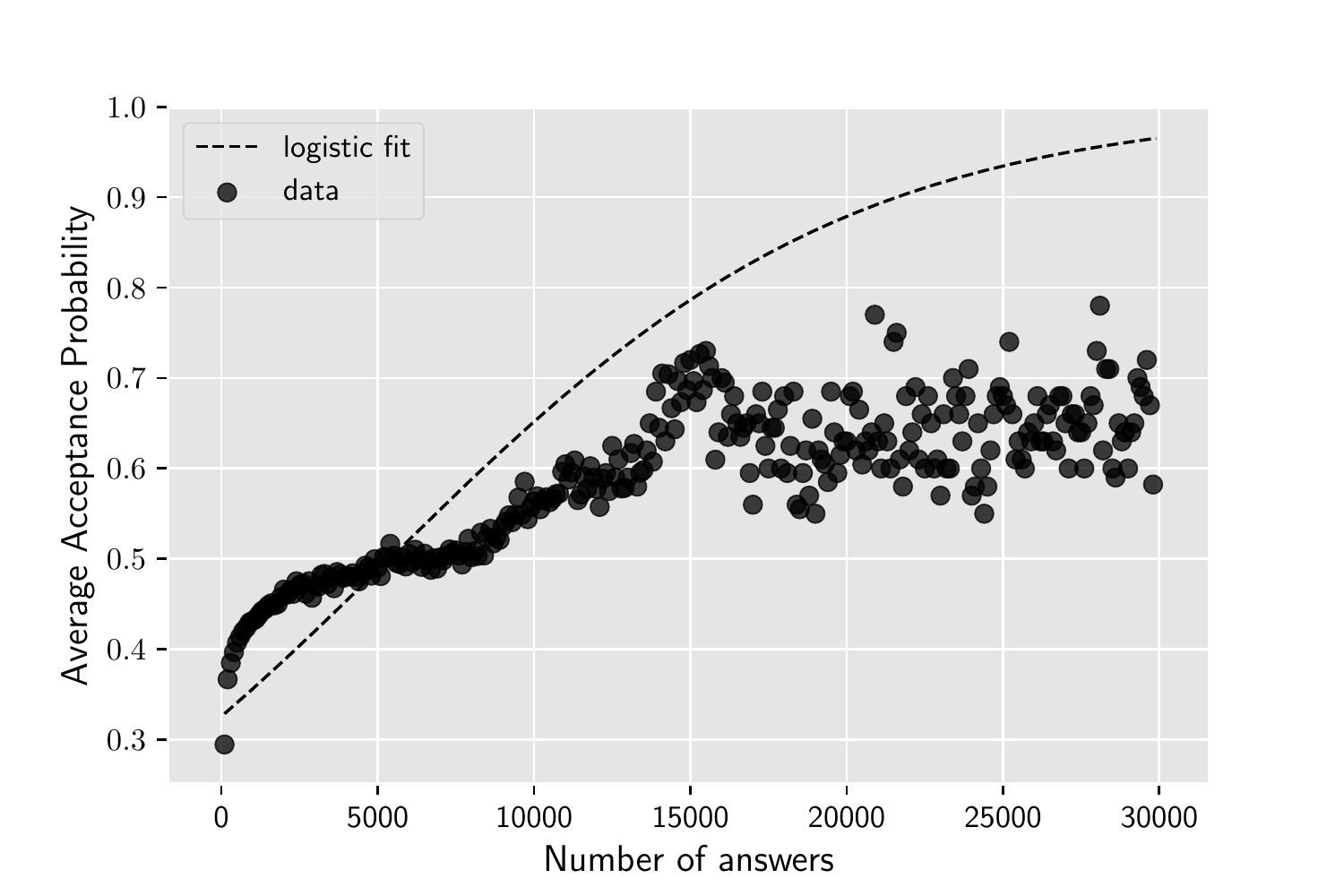}
  \caption{Aggregated Data}
  \label{fig:paradox2a}
\end{subfigure}%
\begin{subfigure}[b]{0.5\textwidth}
  \centering
  \includegraphics[width=\textwidth]{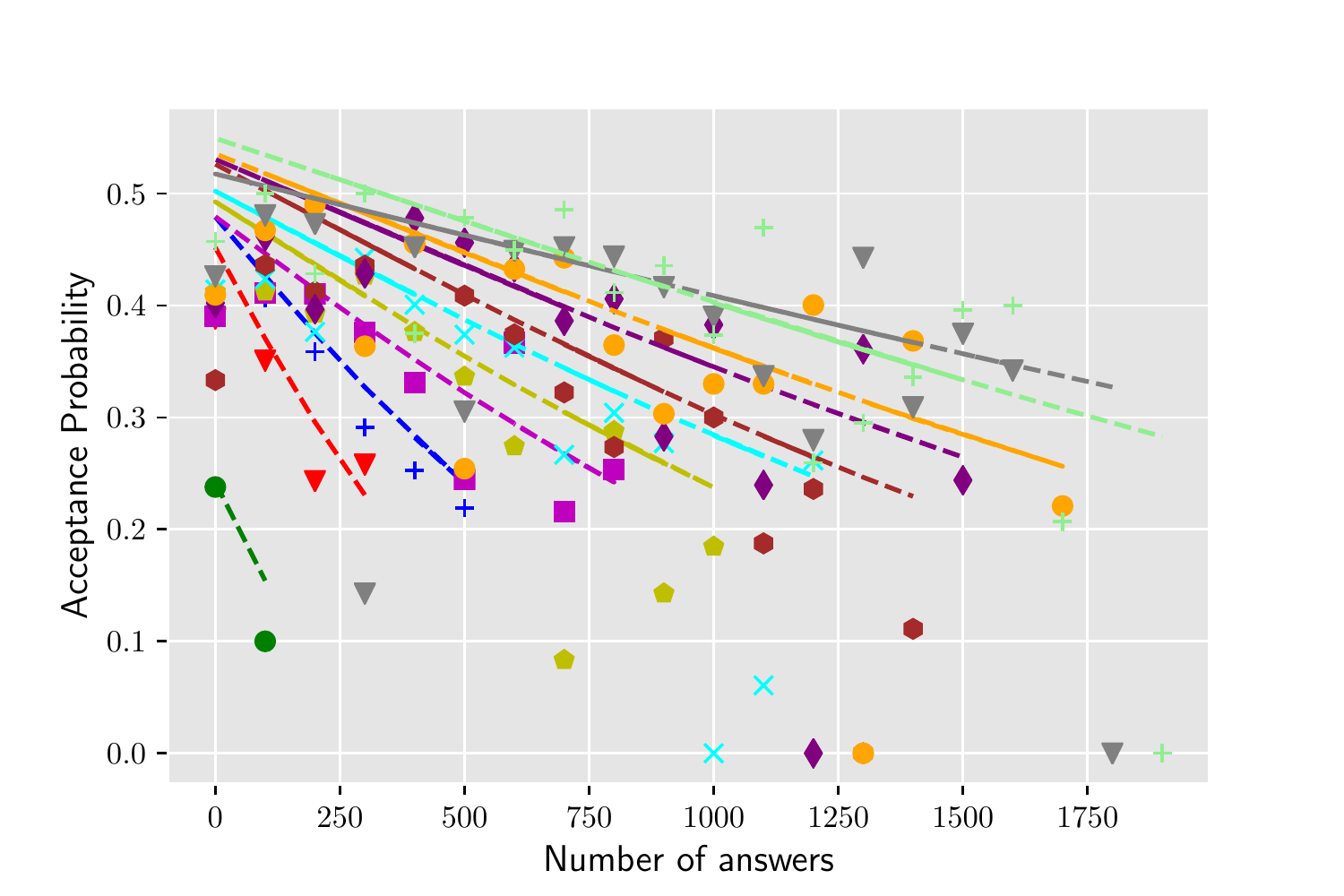}
  \caption{Disaggregated Data}
  \label{fig:paradox2b}
\end{subfigure}
\caption{
Novel Simpson's paradox discovered in Stack Exchange data. Plots show the probability an answer is accepted as best answer as a function of the number of lifetime answers written by user over his or her tenure. (a) Acceptance probability calculated over aggregated data has an upward trend, with answers written by more experienced users (who have already posted more answers) more likely to be accepted as best answers. However, when data is disaggregated by reputation (b), the trend reverses. Among answers written by users with the same reputation (different colors represent reputation bins), those posted by users who had already written more answers are less likely to be accepted as best answers.
}
\label{fig:paradox2}
\end{figure*}

Experience plays an important role in the quality of answers written by users. Stack Exchange veterans, i.e., users who have been active on Stack Exchange for more than six months, post longer, better documented answers, that are also more likely to be accepted as best answers by askers~\cite{Ferrara2017dynamics}. There are several ways to measure experience on Stack Exchange. \textit{Reputation}, according to Stack Exchange, gauges how much the community trusts a user to post good questions and provide useful answers. While reputation can be gained or lost with different actions, a more straightforward measure of experience is user \textit{tenure}, which measures time since the user became active on Stack Exchange, or \textit{Percentile}, normalized rank of a user's tenure.
Alternately, experience can be measured by the \textit{Number of Answers} a user posted during his or her tenure before writing the current answer.

Our method uncovers a novel Simpson's paradox for user experience variables \textit{Reputation} and \textit{Number of Answers}. In the aggregate data, acceptance probability increases as a function of the \emph{Number of Answers} (Fig.~\ref{fig:paradox2a}). This is consistent with our expectations that the more experienced users---who have written more answers over their tenure on Stack Exchange---produce higher quality answers. However, when data is conditioned on \emph{Reputation}, the trend reverses (Fig.~\ref{fig:paradox2b}). In other words, focusing on groups of users with the same reputation, those who have written more answers over their tenure are less likely to have a new answer accepted than the less active answerers.



%
%
%

\subsection{The Origins of Simpson's paradox}
\label{sec:origins}

\begin{table}[b]
  \centering
  \caption{Number of data points in each group}
  \label{tab:sesfreq}
  \resizebox{\columnwidth}{!}{%
  \begin{tabular}{c|cccccccccc}
    \toprule
    \textbf{Session Length}&1&2&3&4&5&6&7&8\\
    \midrule
    \textbf{Data points}& 7.2M & 2.6M & 1.3M & 0.7M & 0.4M & 0.3M & 0.2M & 0.1M\\
    \bottomrule
\end{tabular}
}
\end{table}

\begin{figure*}[th]
  \centering
    \begin{subfigure}[b]{.49\textwidth}
        \includegraphics[width=\textwidth]{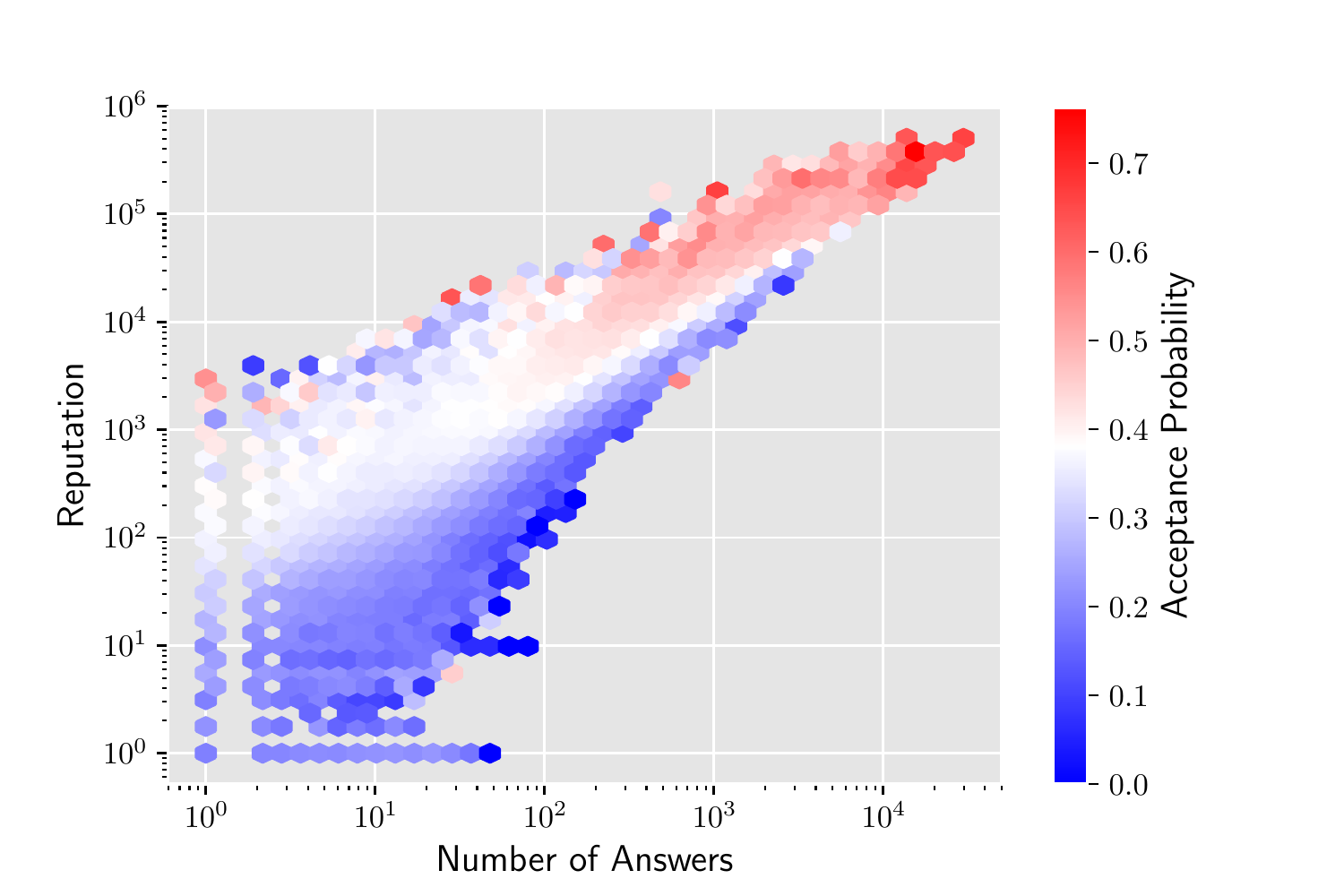}
        \caption{Disaggregated data}  \label{fig:fig3a}
   \end{subfigure}
     \begin{subfigure}[b]{0.49\textwidth}
        \includegraphics[width=\textwidth]{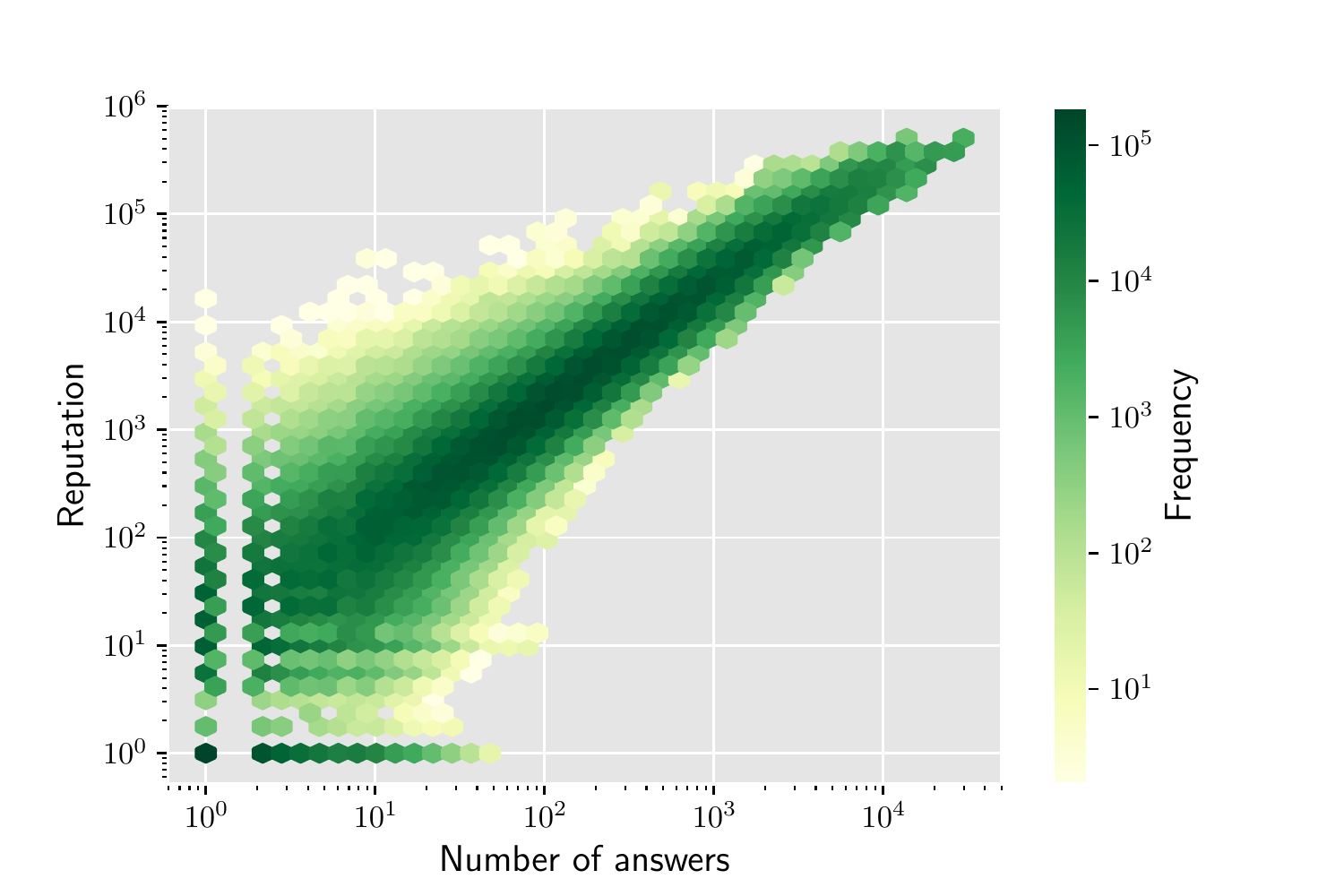}
        \caption{Joint distribution of $X_c$ and $X_p$}
        \label{fig:fig3b}
    \end{subfigure}%
\caption{Analysis of the Simpson's paradox \emph{Reputation} -- \emph{Number of Answers} variable pair. (a) Average acceptance probability as a function of two variables. (b) The distribution of the number of data points contributing to the value of the outcome variable for each pair of variable values. } \label{fig:fig3}
\end{figure*}

To understand why Simpson's paradox occurs in Stack Exchange data, we illustrate the mathematical explanation of Section~\ref{sec:math} with examples from our study.
Consider the paradox for \textit{Answer Position}--\textit{Session Length} Simpson's pair, illustrated in Fig.~\ref{fig:fig1}.
In the disaggregated data, trend lines of acceptance probability for sessions of different length are stacked (Fig.~\ref{fig:fig2}): answers produced during longer sessions are more likely to be accepted than answers produced during shorter sessions. In addition, there are many more shorter sessions than longer ones. Table~\ref{tab:sesfreq} reports the number of sessions of different length. By far, the most common session has length one: users write only one answer during these sessions. Each longer session is about half as common as a session that is one answer shorter.

What happens to the trend in the aggregated data?
When calculating acceptance probability as a function of answer position, all sessions contribute to acceptance probability for the first answer of a session. Sessions of length one dominate the average. When calculating acceptance probability for answers in the second position, sessions of length one do not contribute, and acceptance probability is dominated by data from sessions of length two. Similarly, 
acceptance probability of answers in the third position is dominated by sessions of length three. Survivor bias excludes data from shorter sessions, which also have lower acceptance probability, 
creating an upward trend in acceptance probability.


We back up this intuitive explanation with mathematical analysis of Section~\ref{sec:math}.
Although acceptance probability is decreasing as a function of \emph{Answer Position} for each value of \emph{Session Length} (Fig.~\ref{fig:fig2}), the probability mass of \emph{Session Length} is constantly moving towards larger values as \emph{Answer Position} increases. Notice that as \emph{Answer Position} increments from $a$ to $a+1$, sessions of length $a$ are no longer included (as the minimum session length is now $a+1$). Thus, while \emph{Session Length} has probability mass $\Pr(X_c=a|X_p=a)$ when $X_p=a$, it has probability $\Pr(X_c=a|X_p=a+1) = 0$ at $X_p=a+1$:
\begin{equation}
  \frac{d}{dx_p}\Pr(X_c=a|X_p=x_p)\|_{x_p=a} = -\Pr(X_c=a|X_p=a).
\end{equation}
Meanwhile, for all other values of $X_c$ greater than $a$, the probability mass at $X_p=a+1$ is the same as that at $X_p=a$ (as the number of data points is constant along sessions of same length) but normalized to account for the sessions of length $a$, i.e.,
\begin{equation}
 \Pr(X_c=x_c|X_p=a+1) = \frac{\Pr(X_c=x_c|X_p=a)}{1-\Pr(X_c=a|X_p=a)}.
\end{equation}
The rate of change of these probability masses with respect to $X_p$ is
\begin{multline}
  \frac{d}{dx_p}\Pr(X_c=x_c|X_p=x_p)\|_{x_p=a}  =   \\ \left(\frac{1}{1-\Pr(X_c=a|X_p=a)}-1\right)\Pr(X_c=x_c|X_p=a).
\end{multline}
The probability mass function $\Pr(X_c=x_c|X_p=a)$ decreases for $X_c=a$ corresponding to the smallest value of acceptance probability, while increasing for all other values $X_c > a$. Moreover, the rate of increase of this probability mass is greater than the rate at which the acceptance probability decreases, resulting in an upward trend when the data is aggregated.

%

A similar effect plays out in the \textit{Number of Answers}--\textit{Reputation} Simpson's pair. Figure~\ref{fig:fig3a} shows the heatmap of acceptance probability for different values of the \textit{Number of Answers} written over a user's tenure and user \textit{Reputation}, while Fig.~\ref{fig:fig3b} shows the correlated joint distribution of the two variables.
The figures illustrate the first condition of Simpson's paradox (Eq.~\eqref{eq:SPcond1}): as $X_p$ changes, the distribution of the values of $X_c$ must also change. This dependency can be clearly seen in Fig.~\ref{fig:fig3b}---as $X_p=Number\ of\ Answers$ increases then the distribution of $X_c=Reputation$ shifts to increasing values, which produces the paradox.

In the real world this means that users, who have written more answers are not more likely to have a new answer they write accepted. In fact, among users with same \textit{Reputation}, those who earned this reputation with fewer answers are more likely to have a new answer they write accepted as best answer. This suggests that such users are simply better at answering questions, and that this can be detected early in their tenure on Stack Exchange (while they still have low reputation). 
{Note, however, that an exception to the trend reversal occurs for users with very high reputation. In Stack Exchange, users can gain reputation by ``Answer is marked accepted", ``Answer is voted up", ``Question is voted up", etc. It seems that, high reputation users and low reputation users are different: for high reputation users, experience (number of written answers) is important, while for low reputation users the quality of answers, which may lead to votes, is more important. Analysis of this behavior is beyond the scope of this paper. }

\subsection{Discussion and Implications}

Presence of a Simpson's paradox in data can indicate interesting or surprising patterns~\cite{fabris2000discovering}, and for trends in social data, important behavioral differences within a population. Since social data is often generated by a mixture of subgroups, existence of Simpson's paradox suggests that these subgroups differ systematically and significantly in their behavior. By isolating important subgroups in social data, our method can yield insights into their behaviors.

For example, our method identifies \emph{Session Length} as a conditioning variable for disaggregating data when studying trends in acceptance probability as a function of answer's position within a session. In fact, prior work has identified session length as an important parameter in studies of online performance~\cite{KootiA2017www,Agarwal2017quitting,Singer2016plosone,Ferrara2017dynamics}. Unless activity data is disaggregated into individual sessions---sequences of activity without an extended break---important patterns are obscured. A pervasive pattern in online platforms is user performance deterioration, whereby the quality of a user's contribution decreases over the course of a single session. This deterioration was observed for the quality of answers written on Stack Exchange~\cite{Ferrara2017dynamics}, comments posted on Reddit~\cite{Singer2016plosone}, and the time spent reading posts on Facebook~\cite{KootiA2017www}. Our method automatically identifies position of an action within a session and session length as an important pair of variables describing Stack Exchange.

\begin{figure}[tb]
 \centering
  \includegraphics[width=\columnwidth]{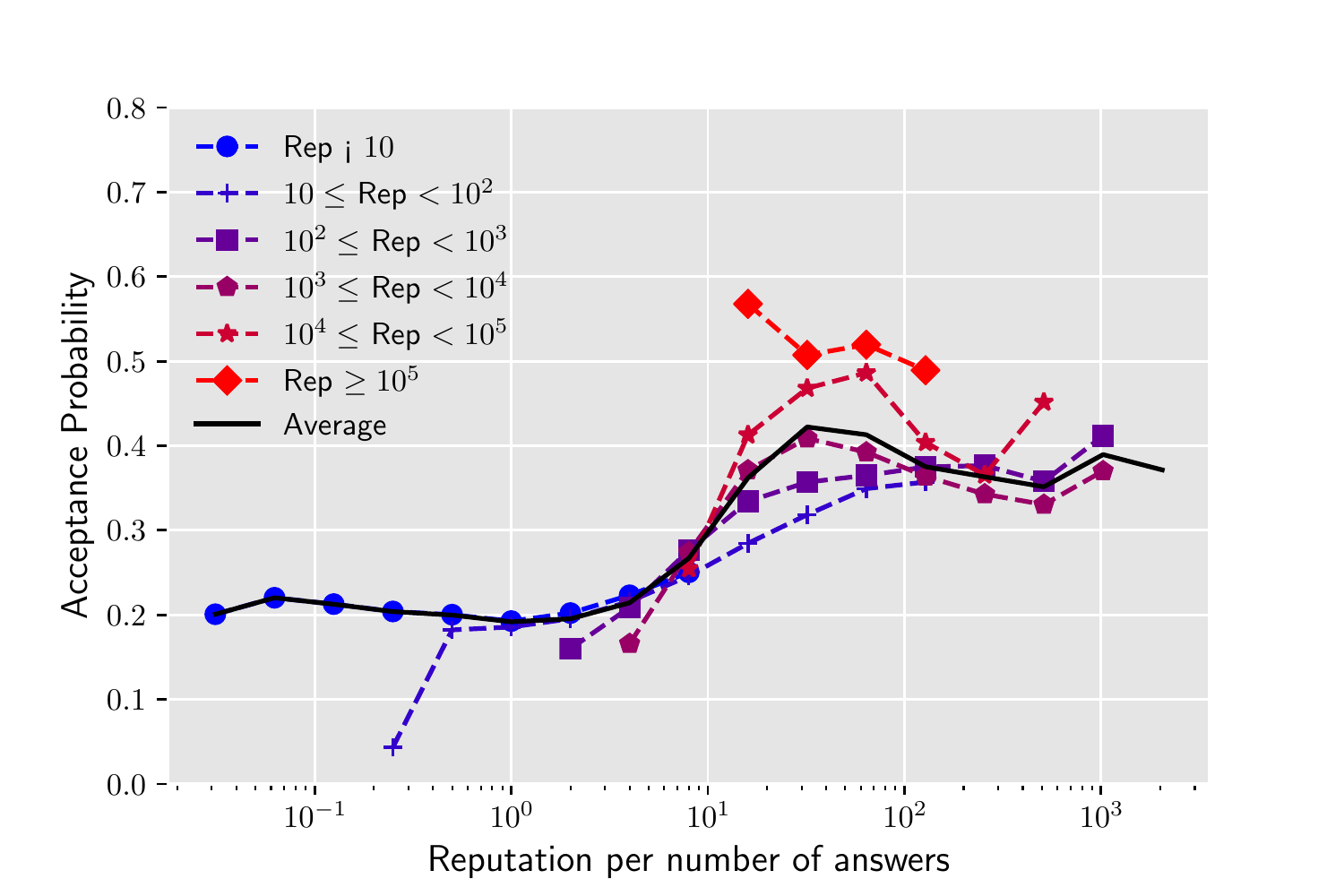}
	\caption{Relationship between acceptance probability and \emph{Reputation Rate}, a new measure of user performance defined as reputation per number of answers users wrote over their entire tenure. Each line represents a subgroup with a different reputation score. The much smaller variance compared to Fig.~\protect\ref{fig:paradox2b} suggests that the new feature is a good proxy of answerer performance.}   \label{fig:var}
\end{figure}

We examine in detail  one novel paradox discovered by our method for the \emph{Reputation}--\emph{Number of Answers} variables.
The trends in Fig.~\ref{fig:paradox2b} suggest that both variables jointly affect acceptance probability. Inspired by this observation, we construct a new variable---\emph{Reputation} /  \emph{Number of Answers}---i.e., \emph{Reputation Rate}. Figure~\protect\ref{fig:var} shows how acceptance probability changes with respect to \emph{Reputation Rate} for different groups of users. There is an strong upward trend, suggesting that answers provided by users with higher \textit{Reputation Rate} are more likely to be accepted. Moreover, while the lines span reputations of an extremely broad range---from one to 100,000---they collapse onto a single curve. This suggests that \emph{Reputation Rate} is a good proxy of user performance.
The remaining paradoxes uncovered by our method could yield similarly interesting insights into user behavior on Stack Exchange.


We also illustrate the difference between our method and linear models that model the outcome variable as a function of both $X_p$ and $X_c$. For such multivariate linear models \cite{Norton2015simpson}, we can fit a model $f_{p,c}(\alpha + \beta X_p + \beta_cX_c)$ to the disaggregated data, and compare the sign of the coefficient $\beta$ to the sign of the linear coefficient of the ``aggregated'' model $f_{p}(\alpha + \beta X_p)$. In our method, we bin the values of $X_c$ and fit separate linear models of the form of Eq.~\eqref{eq:f_pc} in each bin of $X_c$, aggregating by averaging the linear coefficient signs of each model. We claim that our approach has benefits over multivariate linear models which allow it to find Simpson's pairs where multivariate linear models can not. 
First, in multivariate linear models, all subgroups have the same coefficient $\beta$, and intercepts $\alpha + \beta_cX_c$, which vary linearly with $X_c$. In our method, however, each group can have different intercept and coefficient, which makes finding paradox pairs in heterogeneous data more flexible. Indeed this flexibility is necessary --- from our results (Figs.~\ref{fig:fig2} and \ref{fig:paradox2b}) it is clear that the trend parameters $\beta (x_c)$ of the fitted lines vary significantly depending on $x_c$.

\begin{figure*}[th]
  \centering
    \begin{subfigure}[b]{.49\textwidth}
        \includegraphics[width=\textwidth]{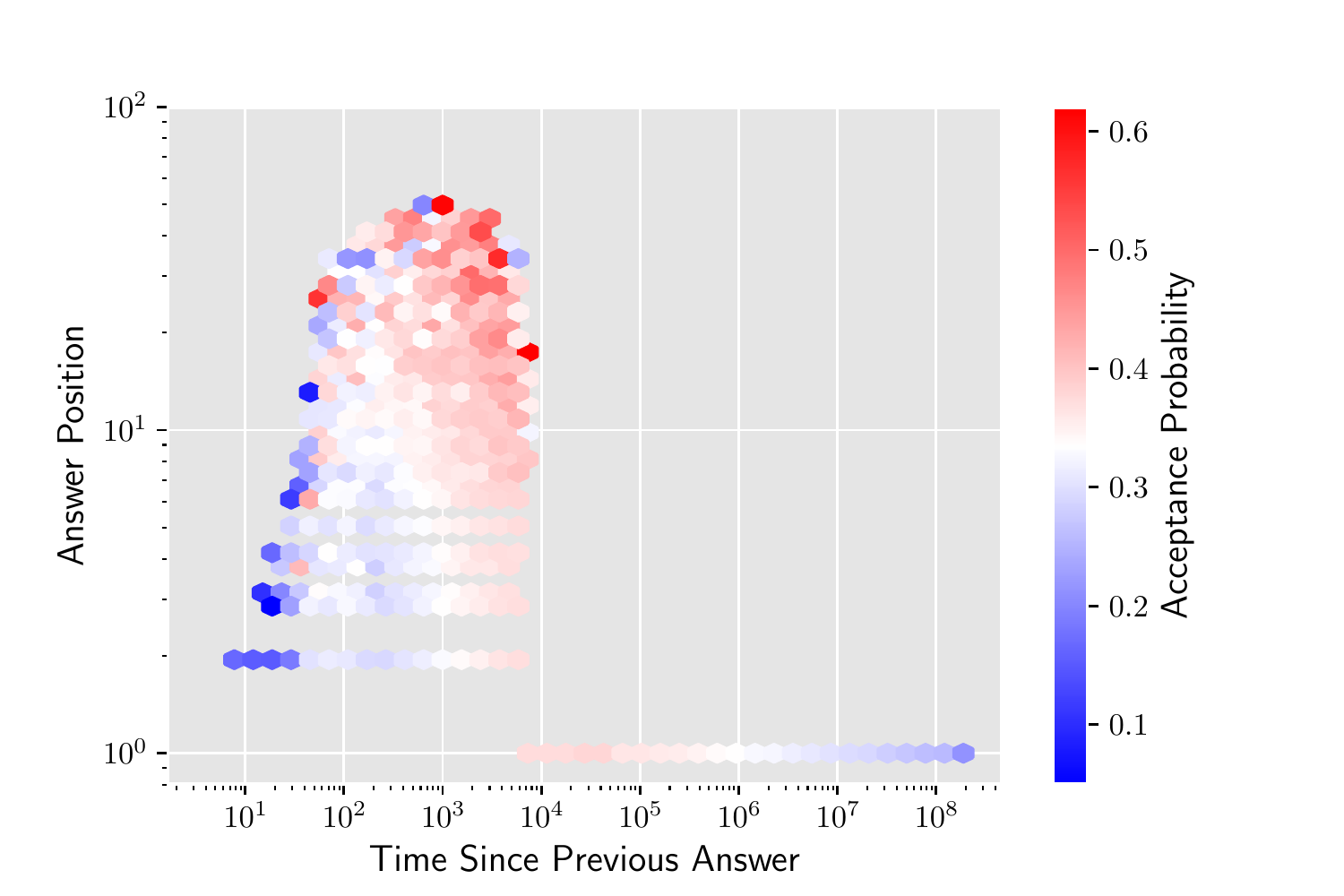}
        \caption{Disaggregated data}  \label{fig:whynotb}
   \end{subfigure}
  \begin{subfigure}[b]{0.49\textwidth}
        \includegraphics[width=\textwidth]{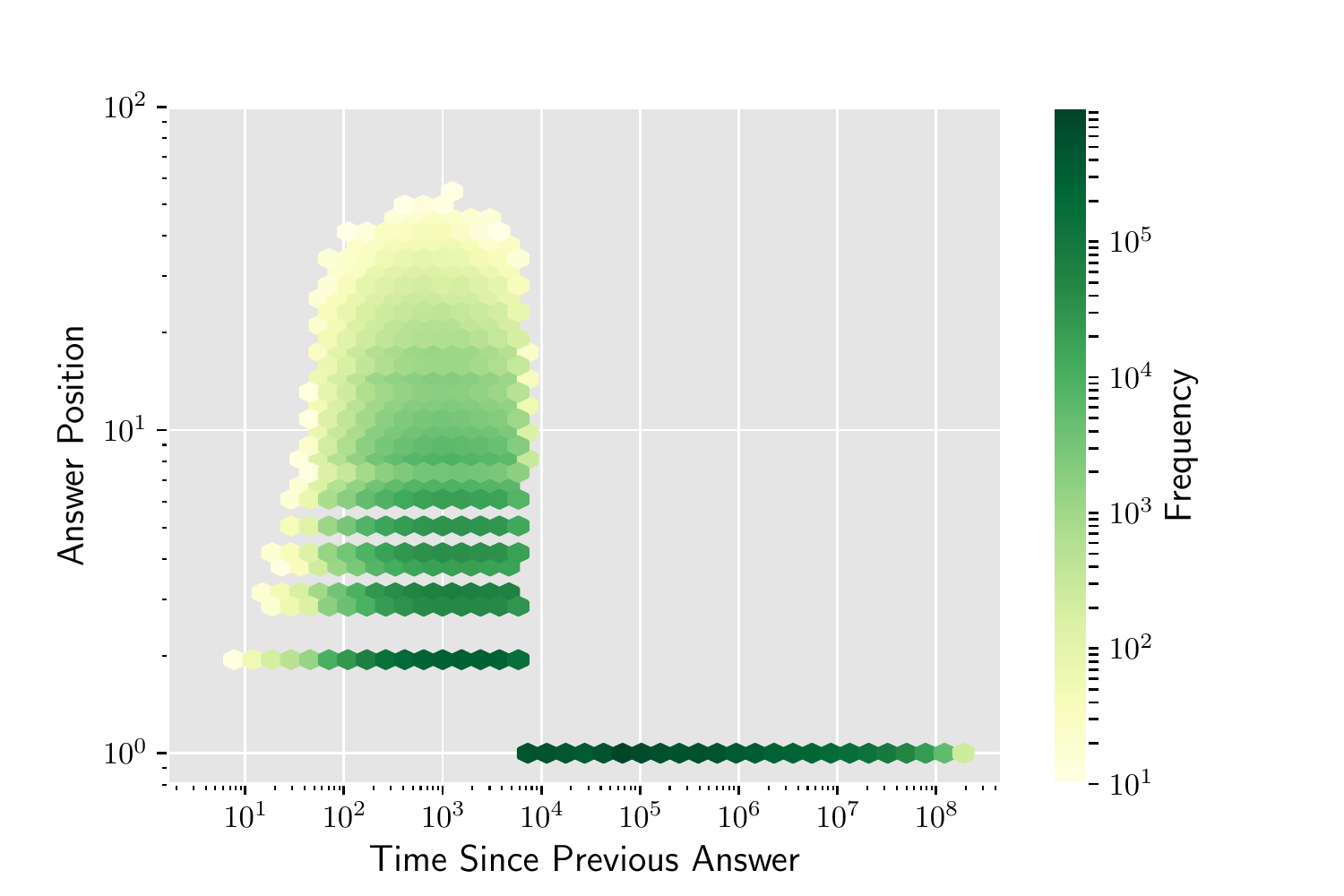}
        \caption{Joint distribution of $X_c$ and $X_p$}
        \label{fig:whynota}
    \end{subfigure}
\caption{A pair which multivariate logistic regression cannot find in the data. (a) Average acceptance probability as a function of Answer Position and Time Since Previous Answer. (b) The distribution of the number of data points contributing to the value of the outcome variable for each pair of variable values. } \label{fig:whynot}
\end{figure*}

Secondly, our method of aggregating by simple averaging of the linear coefficient signs of the subgroups means that trends within each subgroup are weighted equally regardless of how many datapoints are in that subgroup. This is contrary to multivariate linear models, which fit the model parameters based on each datapoint (and so weigh heavily towards values of $X_c$ with many datapoints). To illustrate, we show that our algorithm finds \textit{Time Since Previous Answer - Answer Position} as a Simpson's pair, which a multivariate logistic regression does not. The variable \textit{Answer Position} is the index of the answer a user has completed without an extended (>$100$ minute) break, and so \emph{Answer} $Position=1$ if \textit{Time Since Previous Answer}  $\ge 100$ minutes and \emph{Answer} $Position>1$ if \textit{Time Since Previous Answer} $<100$ minutes. Fig.~\eqref{fig:whynotb} shows that, for \textit{Answer} $Position=1$, the acceptance probability decreases as a function of \textit{Time Since Previous Answer}, possibly because better users take shorter breaks. On the other hand, for other \textit{Answer Positions} the trend is reversed, and acceptance probability increases with \textit{Time Since Previous Answer}, suggesting that in short term, users who take more time to answer questions or take short breaks between questions write answers of higher quality.

Clearly, \textit{Time Since Previous Answer - Answer Position} is an important Simpson's pair, illustrating that time has a beneficial effect on answer quality ar short time scales. even though it is detrimental on the aggregate level. Multivariate logistic regression does not capture this behaviour, as $65\%$  of the probability mass of \textit{Time Since Previous Answer} is for values larger than $100$ minutes, so when fitting $f_{p,c}(\alpha + \beta X_p + \beta_cX_c)$ to the data, it tries to fit a hyperplane, which describes the majority of the data as best as possible, in this case the decreasing trend corresponding to \emph{Answer} $Position=1$.

%

\section{Conclusion}

%
%
%

We presented a method for systematically uncovering instances of Simpson's paradox in data. The method identifies pairs of variables, such that a trend in an outcome as a function of one variable disappears or reverses itself when the same data is disaggregated by conditioning it on the second variable. The disaggregated data corresponds to subgroups within the population generating the data. Our mathematical analysis suggests that Simspon's paradox is caused by both correlations between independent variables in data (Figs.~\ref{fig:fig3b} and \ref{fig:whynota}), as well as differing behaviour of the outcome variable within subgroups, illustrated here by the stacked curves of Figs.~\ref{fig:fig2} and \ref{fig:paradox2b}.
Failure to account for this effect can lead analysis to wrong conclusions about typical behavior of individuals.


We applied our method to real-world data from the question-answering site Stack Exchange. We were specifically interested in uncovering features affecting the probability that an answer written by a user will be accepted by the asker as the best answer to his or her question. We identified eleven relevant features of answers and users. 
Not only did the method confirm an existing paradox, but it also uncovered new instances of Simpson's paradox. 

Our work opens several directions for future work. The proposed algorithm could benefit from a more principled method to bin continuous data and more sophisticated techniques for re-aggregating the intercepts of the curves fitted to disaggregated data. Also, while it appears that conditioning on $X_c$ disaggregates the population into more homogeneous subgroups, we have not used formal methods, such as goodness of fit, to test for better fit of regression models to data. Goodness of fit may also be used to guide data disaggregation strategies. In addition, our method applies to explicitly declared variables, and not to latent variables that may affect data. While these and similar questions remain, our proposed method offers a promising tool for the analysis of heterogeneous social data.

\subsection*{Acknowledgments}
We acknowledge funding support from the James S. McDonnell Foundation,
Air Force Office of Scientific Research (FA9550-17-1-0327), and by the
Army Research Office (W911NF-15-1-0142).


\end{document}